\documentclass[letterpaper]{jpconf}
\usepackage{graphicx}

\begin{document} 

\title{Correlations at intermediate $p_T$ }
\author{Rudolph C. Hwa}
\address{Institute of Theoretical Science and Department of
Physics\\ University of Oregon, Eugene, OR 97403-5203, USA}
\ead{hwa@uoregon.edu}

\begin{abstract}
Correlations among hadrons in jets produced in heavy-ion collisions
are discussed in the framework of the recombination model.  The basic correlation at the
parton level is among the shower partons arising from kinematical
constraint.  The resultant correlation between hadrons at intermediate
$p_T$ is amazingly rich in characteristics.
\end{abstract}
\section{Introduction} 
The study of the correlations at intermediate $p_T$ in heavy-ion
collisions at high energy is important for the understanding of the
interaction between a hard parton and the hot dense medium that it
traverses.  By intermediate $p_T$ we mean the region that stands
between the soft region ($p_T <2$ GeV/c) where the recombination of
thermal partons is most important and the hard region  ($p_T >9$
GeV/c) where the fragmentation of partons is dominant.  Note that this
classification of regions is determined by the modes of hadronization,
which we shall review briefly, rather than by the nature of scattering,
soft or hard.  In the intermediate $p_T$ region the recombination of
the thermal and shower partons is more important than any other
component of hadronization and clearly conveys the medium effect on
hard partons.  And it is in that region where the recent analysis of the
data from RHIC reveals a wealth of information on jet structure.  We
shall examine the properties of correlation in the framework of parton
recombination, which is the only viable hadronization scheme that can
account for the species dependence of the particles produced.  Our
emphasis will be on near-side correlation, which depends mainly on
the correlation among shower partons in a jet.  The away-side
correlation involves other issues besides hadronization and will be the
subject of a future investigation.

\section{Single-particle distribution} 
Before discussing two-particle correlation, it is fitting to review first
the single-particle distribution as determined in the recombination
model \cite{hy1}.  In that model the shower partons in a jet play a
crucial role.  They are semi-hard and can recombine with soft thermal
partons on the one hand, and also with one another on the other
hand.  Their distributions cannot be calculated in perturbative QCD,
but can be determined phenomenologically from the fragmentation
functions (FF), which are themselves determined by fitting the
fragmentation processes in the collisions of simple systems.  In the
framework of parton recombination the shower parton distributions
(SPD) can be extracted from the fragmentation function $D(x)$ by use
of the equation \cite{hy2}
\begin{eqnarray} 
xD^{\pi}_i(x) = \int {dx_1  \over  x_1} {dx_2  \over x_2}
\left\{S^j_i (x_1),\ S^{j'}_i \left({x_2 \over 1-x_1}\right)
\right\} R_{\pi} (x_1, x_2, x) \quad ,
\label{rch1}
\end{eqnarray}
where $i$ specifies the type of hard parton that fragments, $j$ and 
$j^{\prime}$ denotes the types of two partons that recombine, and
$R_{\pi}$ is the recombination function (RF) for the formation of a
pion.  The two SPDs, $S^j_i$ and $S^{j\prime}_i$, are symmetrized in
the order of emission with momentum fractions $x_1$ and $x_2$ [see
(\ref{rch12}) below for the details].  Five such parton distributions have
been determined from five types of $D(x)$ functions \cite{hy2}.  The
RF for pion is \cite{hy3}
\begin{eqnarray} 
R_{\pi} (x_1, x_2, x) = {x_1 x_2  \over  x} \delta  (x_1  +  x_1 - x) ,
\label{rch2}
\end{eqnarray}
and is inferred from pion-induced Drell-Yan process; for proton
formation the details are given also in \cite{hy3}.

In heavy-ion collisions the probability of finding a shower parton $j$
is \cite{hy1}
\begin{eqnarray} 
{\cal S}^j(q) = \xi \sum_i \int dk k
f_i(k) S^j_i (q/k) \ ,
\label{rch3}
\end{eqnarray} 
where $f_i(k)$ is the probability of producing a hard parton of species
$i$ at transverse momentum $k$ \cite{sgf}.  $\xi$ is the average
fraction of the number of hard partons that emerge from the bulk
medium to hadronize in vacuum.

The thermal parton distribution (TPD) is determined by fitting the soft
pion distribution at $p_T <2$ GeV/c by use of the recombination
formula
\begin{eqnarray}
{dN_{\pi}  \over  pdp} = {1 \over p^2}\int {dq_1 \over  q_1}{dq_2
\over  q_2}F_{q\bar{q}} (q_1, q_2) R_{\pi}(q_1, q_2, p) \quad   
\label{rch4}
\end{eqnarray}
where $p_T$ is denoted by $p$, for brevity.  For TPD we use the
factorizable form
\begin{eqnarray}
F_{q\bar{q}} (q_1, q_2) = {\cal T} (q_1) {\cal T} (q_2) \quad  ,  
\label{rch5}
\end{eqnarray}
where
\begin{eqnarray} 
{\cal T}(q) = 
Cqe^{-q/T} .
\label{rch6}
\end{eqnarray}
It is found from the low-$p_T$ data of pions that \cite{hy1}
\begin{eqnarray} 
C = 23.2 \, ({\rm GeV/c})^{-1}, \quad \quad T = 0.317\, {\rm GeV/c}
\label{rch7}
\end{eqnarray}
for central Au-Au collisions.  For non-central collisions the parameters
are given in \cite{ht}.

With these basic quantities specified we can now describe how the pion
distribution can be determined for any $p_T$ by use of the same
equation (\ref{rch4}), but with the two-parton distribution generalized
to include the shower partons.
\begin{eqnarray}
F_{q\bar{q}} (q_1, q_2) = {\cal T} (q_1) {\cal T} (q_2)  + {\cal T}
(q_1) {\cal S} (q_2) + {\cal SS}  (q_1, q_2)
\label{rch8}
\end{eqnarray}
where the last term is written in that way to emphasize that it is not
factorizable, i.e., 
\begin{eqnarray} 
({\cal SS}) (q_1, q_2)=\xi\sum_i
\int dk k f_i(k) \left\{S_i\left({q_1\over
k}\right),S_i\left({q_2\over k-q_1}\right)
\right\}\quad .
\label{rch9}
\end{eqnarray}
In view of (\ref{rch1}) and (\ref{rch4}) it should be clear that the
${\cal SS}$ terms in $F_{q\bar{q}}$ leads to fragmentation 
\begin{eqnarray} 
{dN_{\pi} ^{\cal SS} \over  pdp} = {\xi \over p}\sum_i \int dk 
f_i(k)  D^{\pi}_i\left( {p  \over  k}\right) \quad .
\label{rch10}
\end{eqnarray}

\begin{figure}
\begin{minipage}[b]{2.8in}
\includegraphics[width=1.0\textwidth]{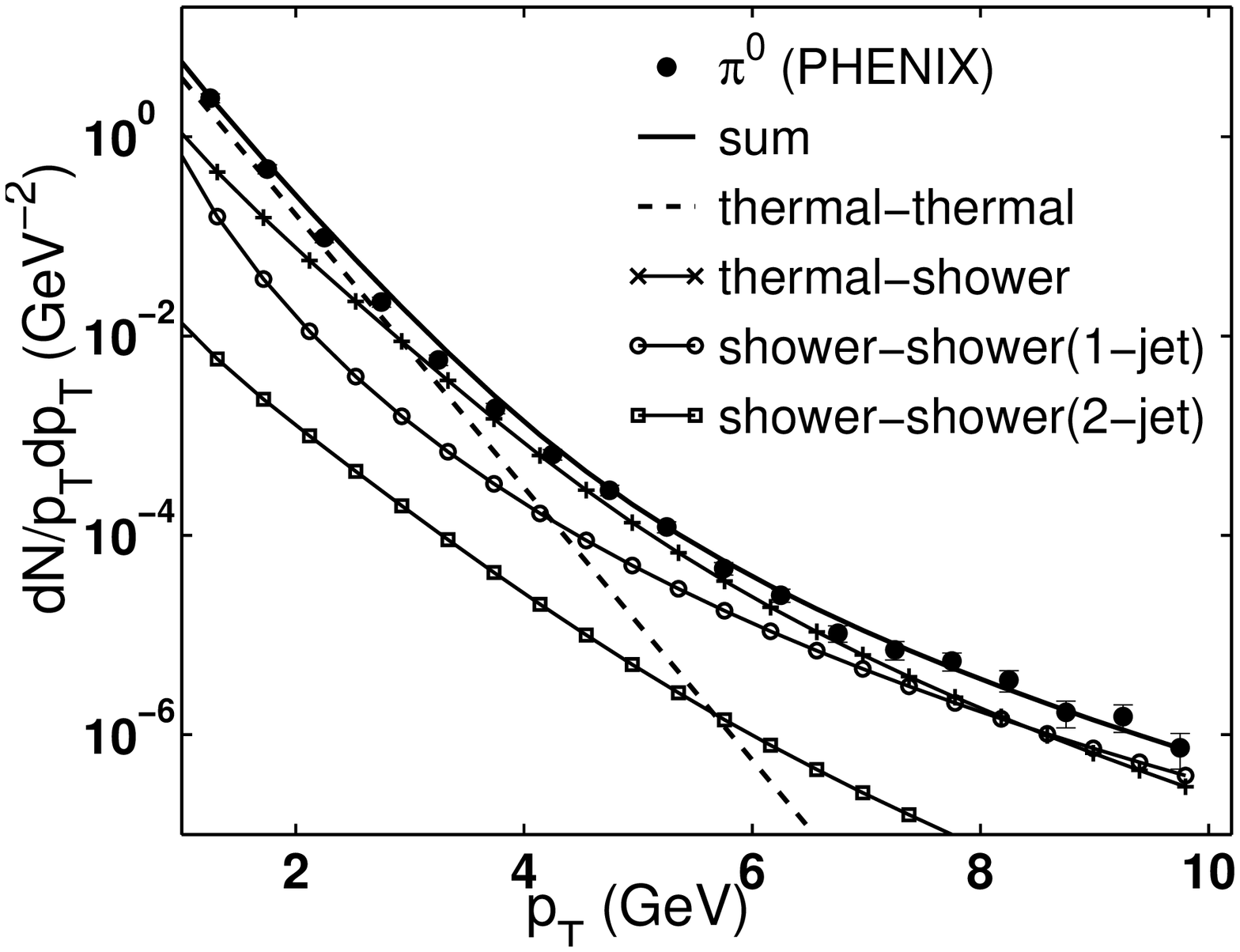}
\caption{Transverse momentum distribution of $\pi^0$ in Au-Au
collisions.  Data in solid circles are from \cite{ssa1}.
}
\end{minipage}
\hspace*{.5in}
\begin{minipage}[b]{2.8in}
\includegraphics[width=1.0\textwidth]{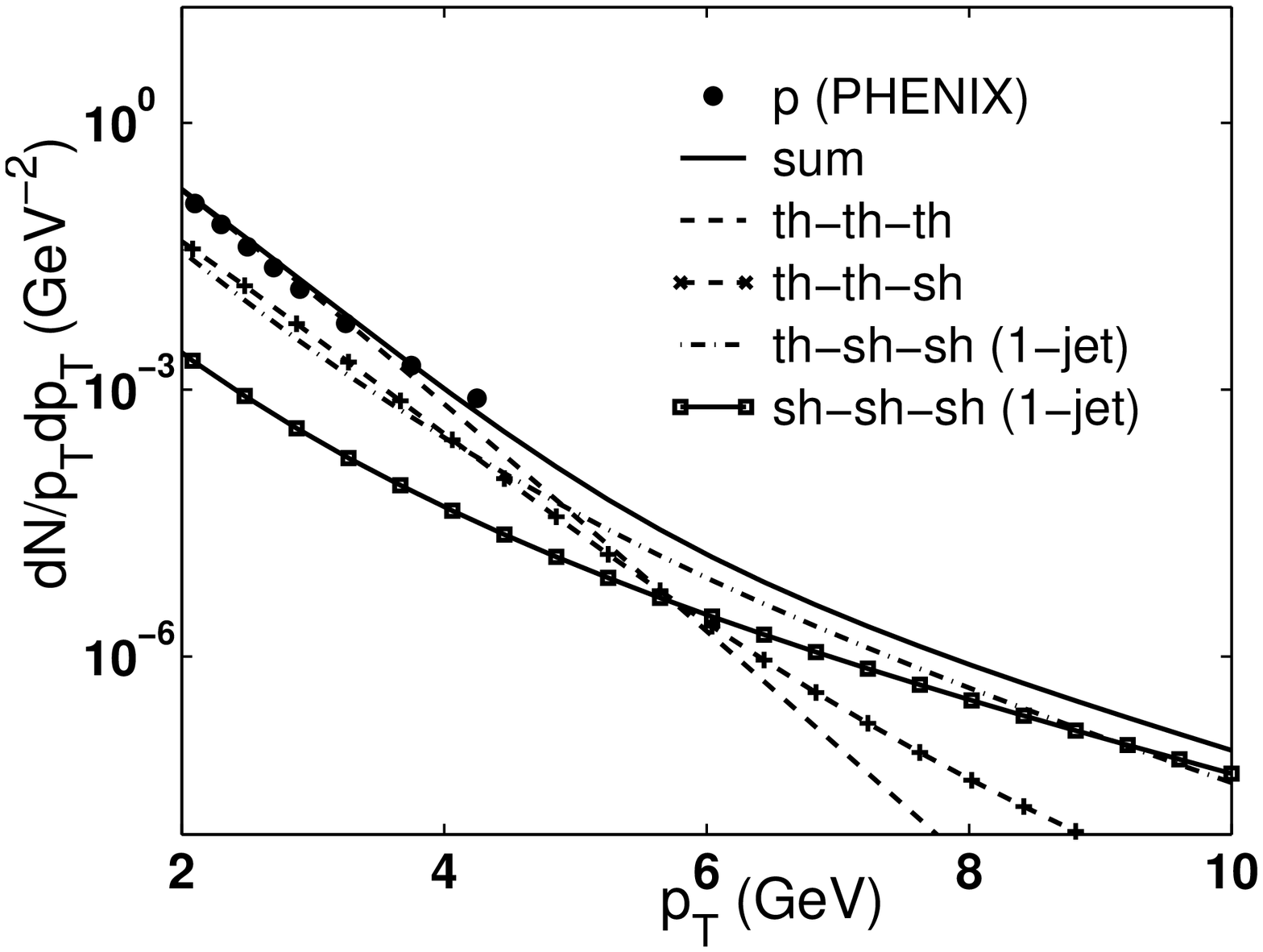}
\caption{Transverse momentum distribution of proton in Au-Au
collisions.  Data in solid circles are from \cite{ssa1}.
}
\end{minipage}
\end{figure}

What is new is the ${\cal TS}$ term in (\ref{rch8}); it dominates in the
intermediate $p_T$  region, as evidenced in Fig.\ 1, in which the overall
normalization is adjusted to fit the data \cite{ssa1} by letting $\xi$ be
0.07.  The shape of the $p_T$ dependence is a prediction of the model. 
In that figure the shower-shower (2 jet) line corresponds to the
recombination of shower partons arising from two different jets, and
should be ignored for collisions at RHIC energies.  The dominance of
${\cal TS}$ recombination in the $3 < p_T <9$ GeV/c region cannot
be reproduced by fragmentation even if the FF used is
medium-modified because the momentum fraction $x$ in the FF
requires the parton momentum to be greater than the pion
momentum $p$, whereas the RF requires the coalescing parton
momenta to be less than $p$.  Since the parton momenta are damped
by a power law, the latter process always wins.  

The contrast between the two processes of hadronization becomes
even more pronounced in the case of proton production.  Since three
quarks recombine to form a proton, the average parton momentum is
$p/3$, so they are even more abundantly available.  To form a proton
by fragmentation, one pays a heavy penalty to produce a high $k$
parton, and then pays an even heavier penalty to require that it
fragments into a proton, the FF for which is an order of magnitude
smaller than $D^{\pi}$.  This is why the $p/\pi$ ratio can be high in
the recombination model but very small in the fragmentation model. 
The production of proton in central AuAu collisions has been calculated in the recombination model where $\cal TTS$ and $\cal TSS$ components have been found to be more important than $\cal SSS$ component (i.e., fragmentation) for $p_T<9$ GeV/c \cite{hy1}. This is shown in Fig.\ 2, where the data \cite{ssa1} exist only up to 4 GeV/c. But that is enough to exhibit the large $p/\pi$ ratio \cite{ssa2}, as shown in Fig.\ 3, where the dashed line takes the proton mass into account at low $p_T$ \cite{hy1}. Actually, the large $p/\pi$ ratio was obtained in an earlier paper on recombination \cite{hy3.5} even before the shower parton distributions were obtained. The parton distributions there were inferred from the pion distribution.  Two other groups have also obtained
similar results using recombination/coalescence model \cite{gre,rjf}.  

\begin{figure}
\begin{minipage}[b]{2.8in}
\includegraphics[width=1.0\textwidth]{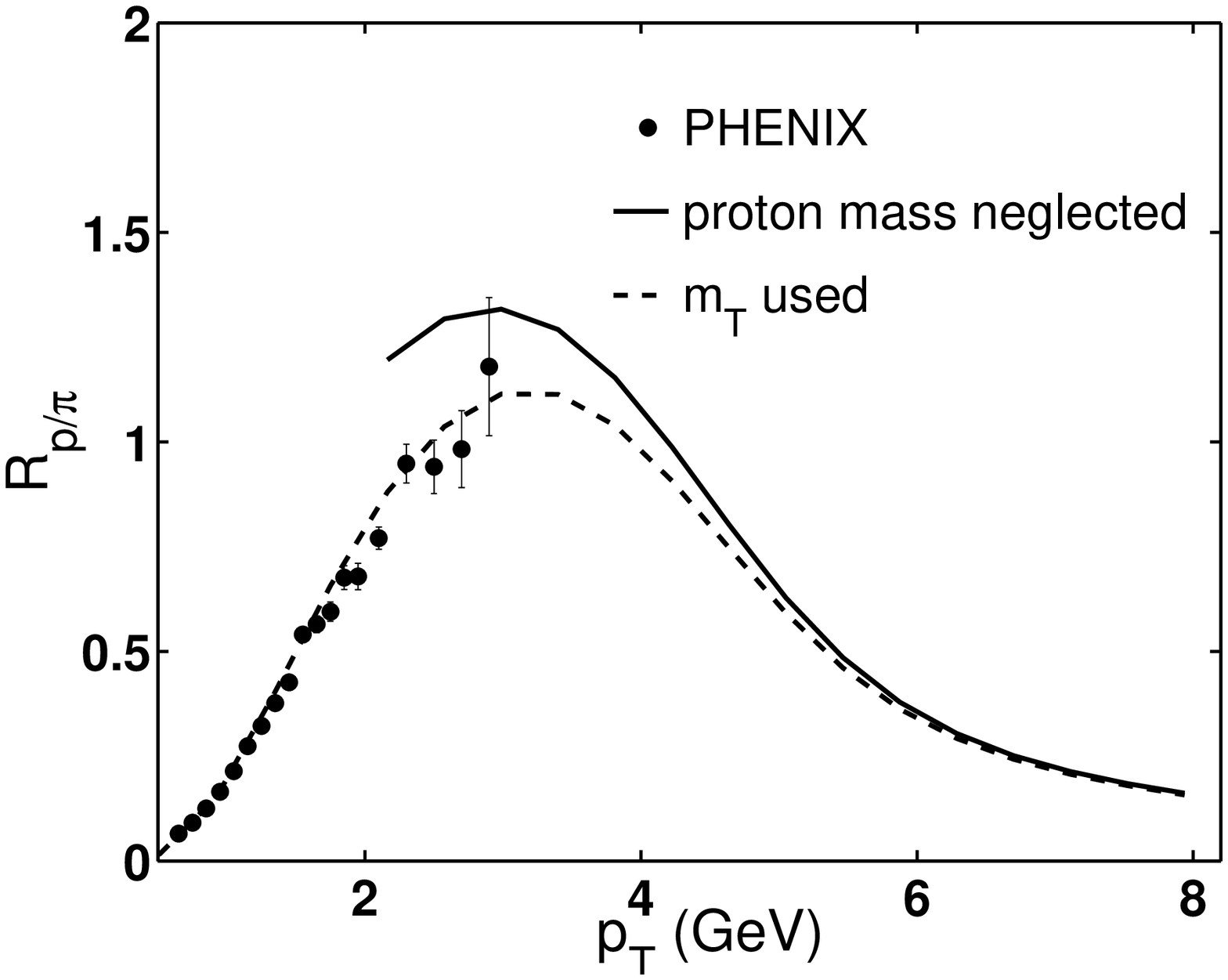}
\caption{Comparison of calculated $p/\pi$ ratio with data from
\cite{ssa2} on  AuAu collisions.}
\end{minipage}
\hspace*{.5in}
\begin{minipage}[b]{2.8in}
\includegraphics[width=1.0\textwidth]{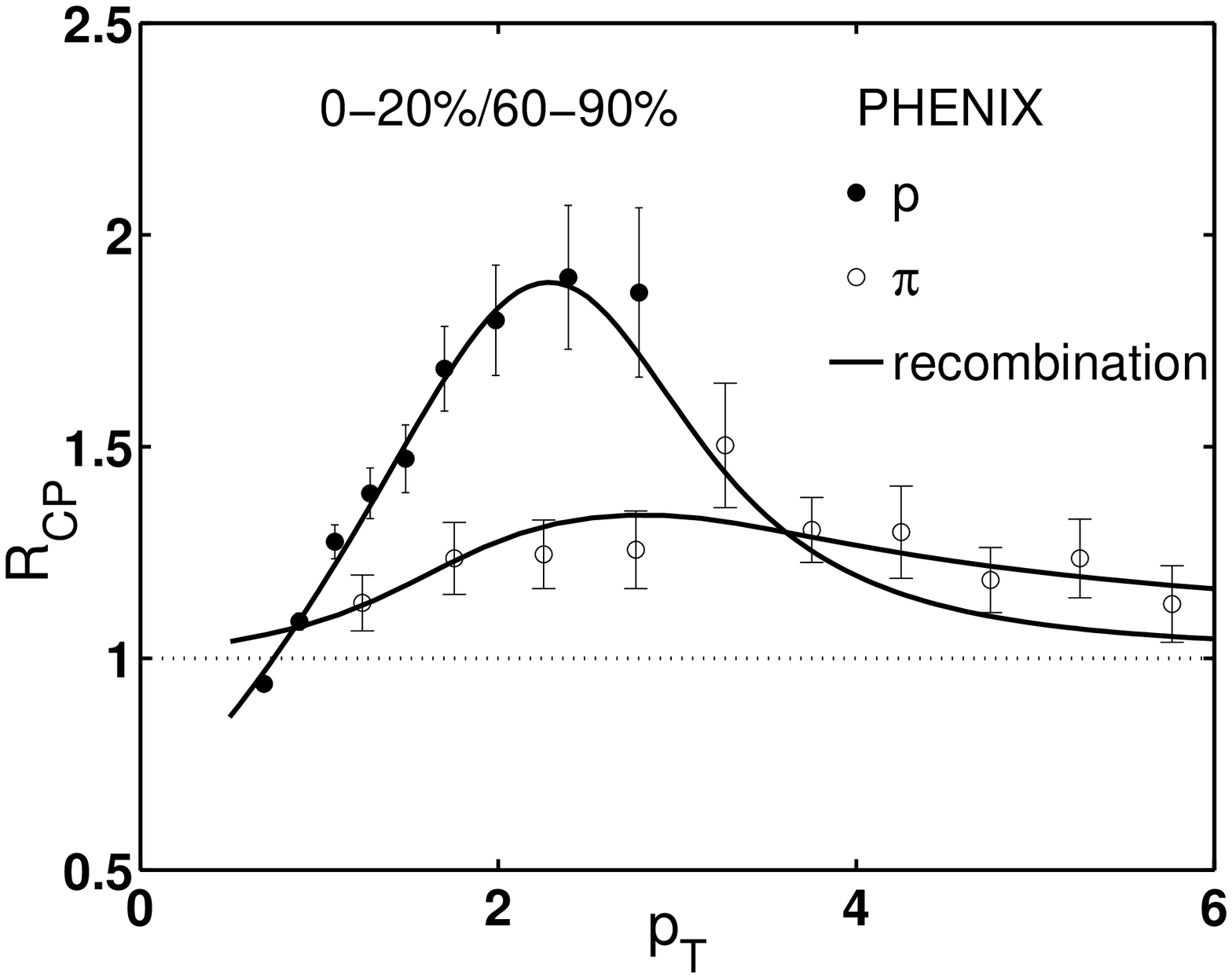}
\caption{$R_{CP}$ for proton and pion in d-Au collisions. The data are from \cite{fm}.}
\end{minipage}
\end{figure}

The Cronin effect has for thirty years been referred to as the
manifestation of $k_T$ broadening by multiple scattering in the initial
state of pA collisions.  That relationship does not take into account of
the fact that the experimental $p_T$ spectrum in $p + A \to h+X$
depends on $A$ as $A^{\alpha _h}$, where $\alpha_p > \alpha_{\pi}$
\cite{cr}.  If the effect of the nuclear medium on hard scattering is
before fragmentation, then the exponent $\alpha_h$ should be
independent of whether the hadron $h$ is a pion or a proton.  In
reality, not only is $\alpha_p >  \alpha_{\pi}$  experimentally, the
FF for proton $D^p$ is much smaller than that for pion, $D^{\pi}$,
by roughly an order of magnitude.  This failure in interpreting the
data has been corrected by use of parton recombination as the
hadronization mechanism.  We have studied the  production of hadrons (pion and proton) at intermediate $p_T$ in d-Au collisions at all centralities in the recombination model \cite{hy4}. Fig.\ 4 shows our results on $R_{CP}$ for pion and proton. Evidently, we obtain $R^p_{CP}>
R^{\pi}_{CP}$ in the range $1<p_T<3$ GeV/c, in good agreement with the data \cite{fm}. This result may be regarded as the strongest support for the recombination model, since no other approaches have indicated the possibility of attaining the same.

One last feature of single-particle distributions that we choose to
mention here is the suppression of $R_{CP}$ in forward production. 
BRAHMS data on d-Au collision \cite{iar} have shown that $R_{CP}$ is
as low as 0.5 at $\eta = 3.2$.  Such a suppression of central production
has been interpreted as suggestive evidence for color glass condensate,
since at large $\eta$ the small-$x$ nuclear partons are presumed to be
important and their high density there reveals saturation physics.  We
have, however, calculated the $p_T$  spectra at large $\eta$ in the
recombination model without incorporating any exotic physics, and
found results in agreement with the data \cite{iar,hyf}.  Our input is
the data on $dN_{ch}/d\eta$ which decreases with increasing $\eta$
much more rapidly for central d-Au collisions than for peripheral
collisions.  Since $dN_{ch}/d\eta$ is dominated by soft partons, the
${\cal TS}$ recombination results in the corresponding decrease of the
hadron distributions at large $\eta$.  There is no change of the
underlying physics as $\eta$ is carried from backward to forward
direction.

\section{Parton and hadron correlations in jets}
Having established some degree of reliability of parton recombination
in the treatment of single-particle distributions, we now consider
correlation of particles in jets at high $p_T$.  There are two ways to
study correlation:  one is to use trigger particles to select events in
which the associated particles reveal near-side and away-side
characteristics; the other is to treat the two particles on equal footing
and study the two-particle correlation function.  Both approaches have
been adopted by different groups within the STAR collaboration.  We
have some recent results on both that can be reported here, starting
with the latter.

In general, the two-particle correlation is defined by
\begin{eqnarray}
C_2 (1,2) = \rho_2 (1,2) - \rho_1 (1)\rho_1 (2) \quad .
\label{rch11}
\end{eqnarray}
where $\rho_2 (1,2)$ is the two-particle distribution, and $\rho_1
(1)$ is the one-particle distribution, which for pion production at
intermediate $p_T$ is given by (\ref{rch4}) in the recombination
model.  It is, however, more appropriate to discuss first, not particle
correlation, but parton correlation in a jet.  Consider a hard parton
with a fixed momentum $k$ in vacuum, as in $e^+e^-$ annihilation.
Since we shall discuss the correlation in terms of momentum fractions
$x_i$, it does not matter what $k$ is so long as it is high enough.  Thus
for a single shower parton, we have $\rho_1 (1) = S^j_i(x_1)$.

For two shower partons we have
\begin{eqnarray}
\rho_1 (1, 2)=\left\{S^j_i (x_1),\ S^{j'}_i \left({x_2\over
1-x_1}\right)\right\} = {1\over  2} \left[S^j_i (x_1) S^{j'}_i
\left({x_2\over 1-x_1}\right) + S^j_i
\left({x_1\over 1-x_2}\right)S^{j'}_i (x_2)\right] 
\label{rch12}
\end{eqnarray}
which guarantees that $x_1+ x_2 \leq 1$ and symmetrizes the order of
emission.  Evidently, the two partons are correlated by virtue of the
form in (\ref{rch12}).  If we define the normalized distribution by the
ratio
\begin{eqnarray}
r_2 (1,2) = {\rho_2 (1,2) \over  \rho_1 (1)\rho_1
(2)}  \quad ,
\label{rch13}
\end{eqnarray}
then the result of our calculation for $r_2 (1,2)$ is shown in Fig.\ 5
\cite{ht}.  It is clear that there is correlation for almost all $x_1$ and
$x_2$ except when they are very small, where $r_2 (1,2) \approx 1$.

\begin{figure}
\begin{minipage}[b]{2.8in}
\includegraphics[width=1.0\textwidth]{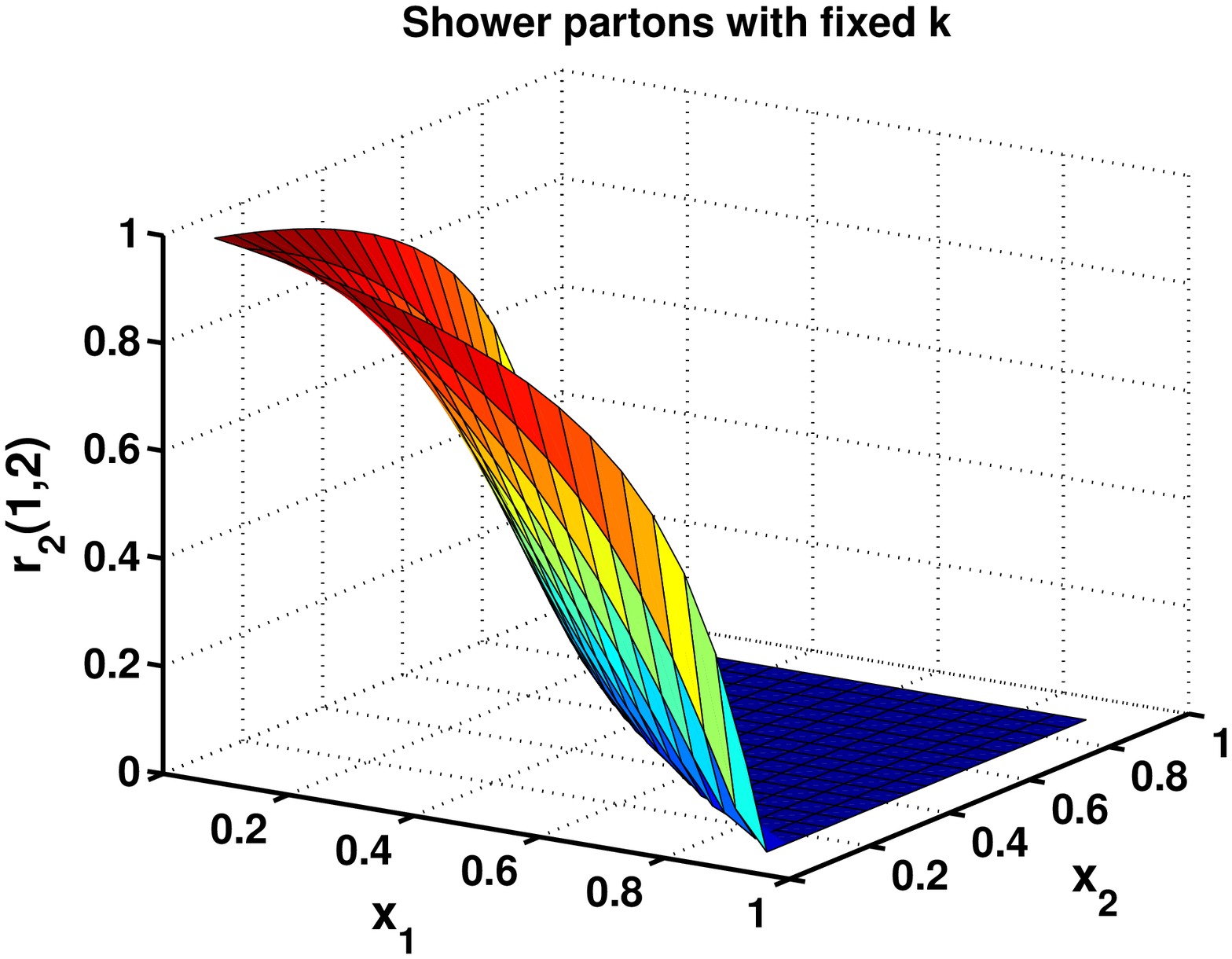}
\caption{The ratio $r_2 (1,2)$ in terms of the momentum fractions of two
shower partons in a hard gluon jet.
}
\end{minipage}
\hspace*{.3in}
\begin{minipage}[b]{3.2in}
\includegraphics[width=1.0\textwidth]{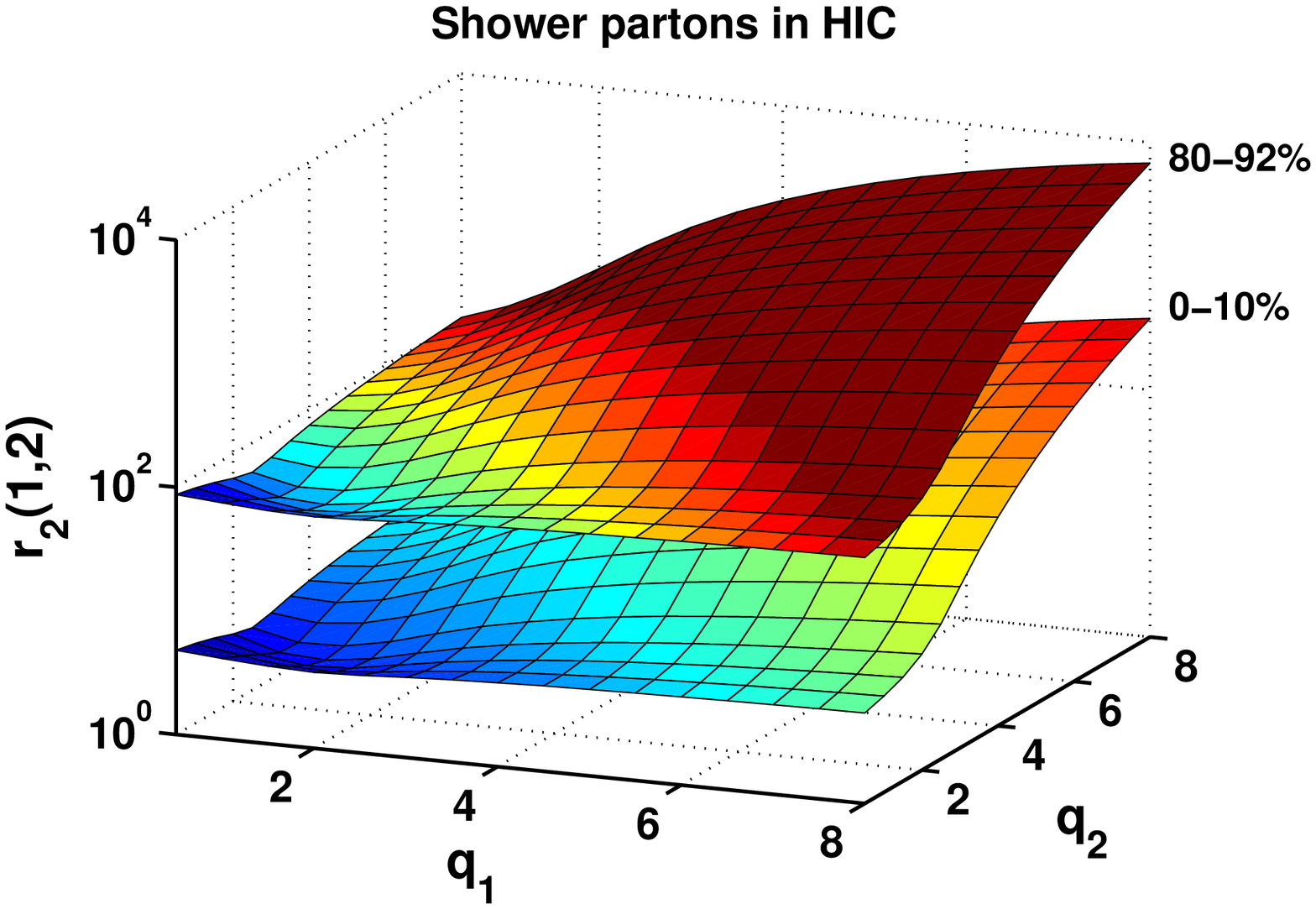}
\caption{The ratio $r_2 (1,2)$ in terms of the momenta of two
shower partons in heavy-ion collisions for two extreme centralities.
}
\end{minipage}
\end{figure}

Now, consider two shower partons in a jet in heavy-ion collisions.  In
that case the hard parton momentum $k$ is not fixed, so the
corresponding  $\rho_1$ and $\rho_2$ involve integrals over $k$,
i.e., 
\begin{eqnarray}
\rho_1 (1) = {\cal S} (q_1)
  \quad  ,\quad
\rho_2 (1,2)=  ({\cal SS})^{jj^{\prime}} (q_1, q_2)
  \quad ,
\label{rch14}
\end{eqnarray}
where ${\cal S}$ and ${\cal SS}$ are given in (\ref{rch3}) and
(\ref{rch9}).  The calculated results for $r_2 (1,2)$ in that case are
shown in Fig.\ 6 for both central and peripheral collisions \cite{ht}. 
They become very large at large $q_1$ and $q_2$ because each
$\rho_1$ is power damped at large $k$, as is $\rho_2$.  The drastic
difference between Figs.\ 5 and 6 underscores the effect of hard
scattering in heavy-ion collision even when the only correlation in the
problem is the same in both cases.

The correlation between pions in jets is far more complicated to
calculate because of the many ways that partons can recombine.  The
two-pion distribution is 
\begin{eqnarray} 
\rho_2 (1,2)= {dN_{\pi_1 \pi_2}  \over  p_1 p_2 dp_1d p_2} =  {1
\over (p_1 p_2)^2}
\int\left(\prod_i {dq_i  \over  q_i}\right) F_4
(q_1, q_2,q_3, q_4) R(q_1, q_3, p_1)
R(q_2, q_4, p_2) .
\label{rch15}
\end{eqnarray}
where
\begin{eqnarray}
F_4 = ({\cal TT} + {\cal ST} + {\cal SS}) _{13} ({\cal TT} + {\cal
ST} + {\cal SS}) _{24} \quad .
\label{rch16}
\end{eqnarray}
While many parts of $F_4$ are factorizable, and therefore make no
contribution to $C_2 (1,2)$, there are non-factorizable parts that
involve at least one ${\cal S}$ in each of $(\cdots) _{ 13}$ and
$(\cdots) _{24}$, the most important example of which is $({\cal
ST})_{13}({\cal ST})_{24}$.  The two ${\cal S}$ terms in that
component, involving the shower partons $S(q_1)$ and $S(q_2)$ are
correlated because they are in  the same jet.  Using (\ref{rch15}) and
(\ref{rch4}) in (\ref{rch11}), we obtain
$C_2 (1,2)$ which is shown in Fig.\ 7.  There is not too much
difference in the shapes of $C_2$ for the central and peripheral cases
for most of $p_1$ and $p_2$, except when the momenta are small
where $C_2$ becomes negative for central collisions and therefore
cannot be exhibited in the log plot \cite{ht}.  The rapid decrease at
large momenta is due to the power-law damping of $f_i(k)$ in both
$\rho _1$ and $\rho _2$.  
In the lower part of the intermediate $p_T$ region various components of $C_2$ become negative, as shown in Fig.\ 8. That occurs because of the competition for the shower parton momenta in a jet, when the hard parton momentum $k$ can be low enough to avoid the severe suppression of $f_i(k)$. Since the correlation of the shower partons is negative, as we have seen in Fig.\ 5, it is not surprising that $C_2$ for the hadrons also becomes negative when $p_1$ and $p_2$ are not too high.

\begin{figure}
\begin{minipage}[b]{3.2in}
\includegraphics[width=1.0\textwidth]{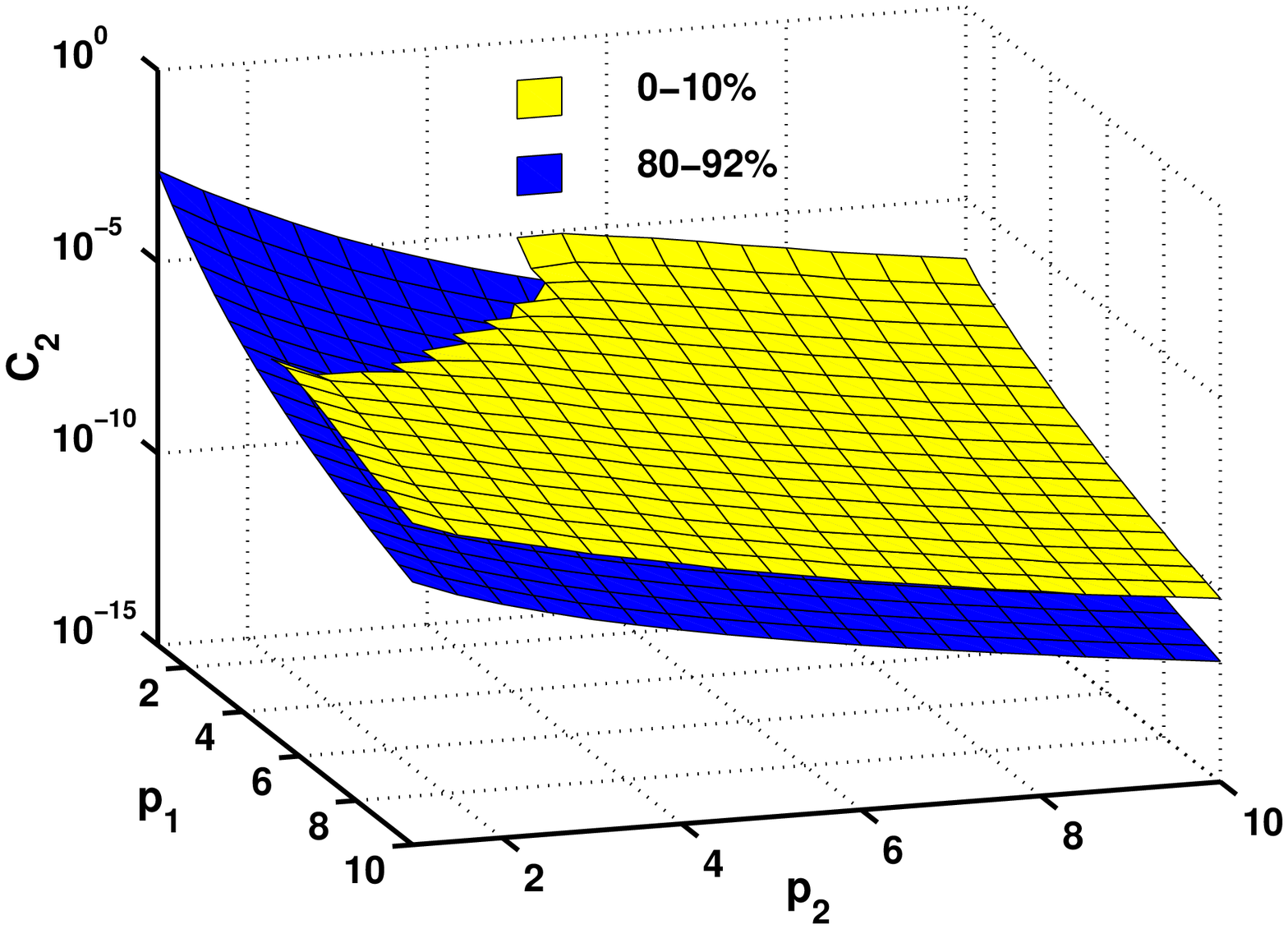}
\caption{The correlation function $C_2$ 
for two centralities when they are positive. At low values of $p_1$ and $p_2$, $C_2$ becomes negative for central collisions.}
\end{minipage}
\hspace*{.1in}
\begin{minipage}[b]{3in}
\includegraphics[width=1.0\textwidth]{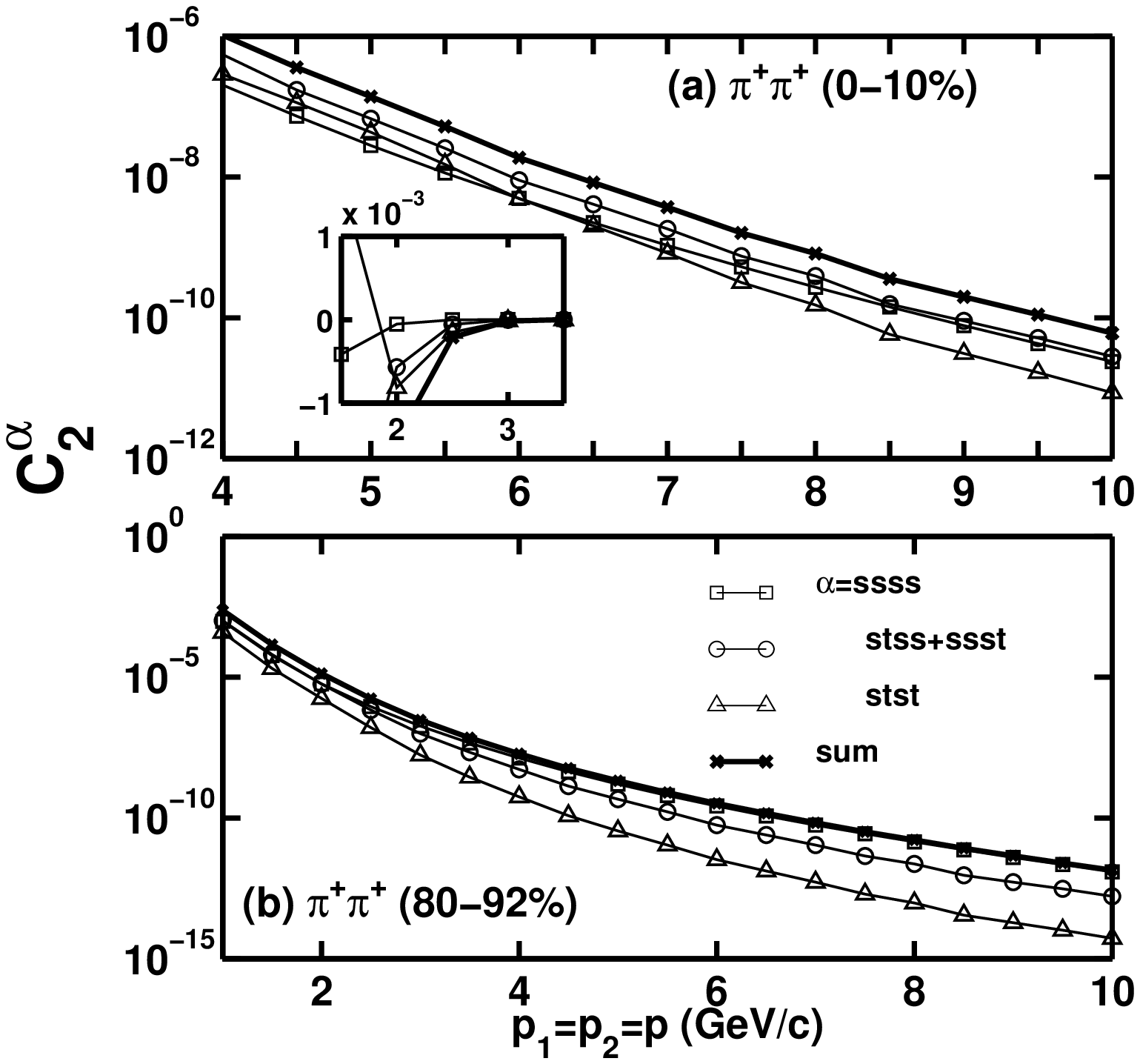}
\caption{The correlation function $C_2$ 
for (a) central and (b) peripheral collisions along the diagonal where
$p_1 = p_2$.}
\end{minipage}
\end{figure}

To compensate for that suppression at high $p_T$ let us
define
\begin{eqnarray}
G_2 (1,2) = {C_2 (1,2) \over  \left[\rho_1 (1)\rho_1
(2)\right]^{1/2}}  \quad ,
\label{rch17}
\end{eqnarray}
whose dependence on $p_1$ and $p_2$ can now be shown in  linear
plots. Fig.\ 9(a)  shows that for central collisions $G_2$ becomes
negative for $p_1$ and/or $p_2 \stackrel{<}{\sim} 4$ GeV/c.  The
ratio, $R^{G_2}_{CP}$, of $G_2$ for the two extreme centralities
exhibits a minimum at $p_1 \approx p_2 \approx 2$ GeV/c, as shown
in Fig.\ 9(b) \cite{ht}.  Data on that ratio, thus far not analyzed, would be
able to provide information on whether there exist  any dynamical
correlations that we have not incorporated in our calculation.

So far we have only considered the correlations in the momentum
variables $p_1$ and $p_2$ of $C_2 (1,2)$.  We can also study the
autocorrelation in $\Delta\eta $ and $\Delta \phi$, for which there are
data at low $p_T$ \cite{rt}.  For such studies we need information on
the angular distribution of shower partons.  With that goal in mind we
turn next to the investigation of correlations with trigger particles
selected to serve as reference.

\begin{figure}
\begin{minipage}[b]{3.0in}
\includegraphics[width=1.0\textwidth]{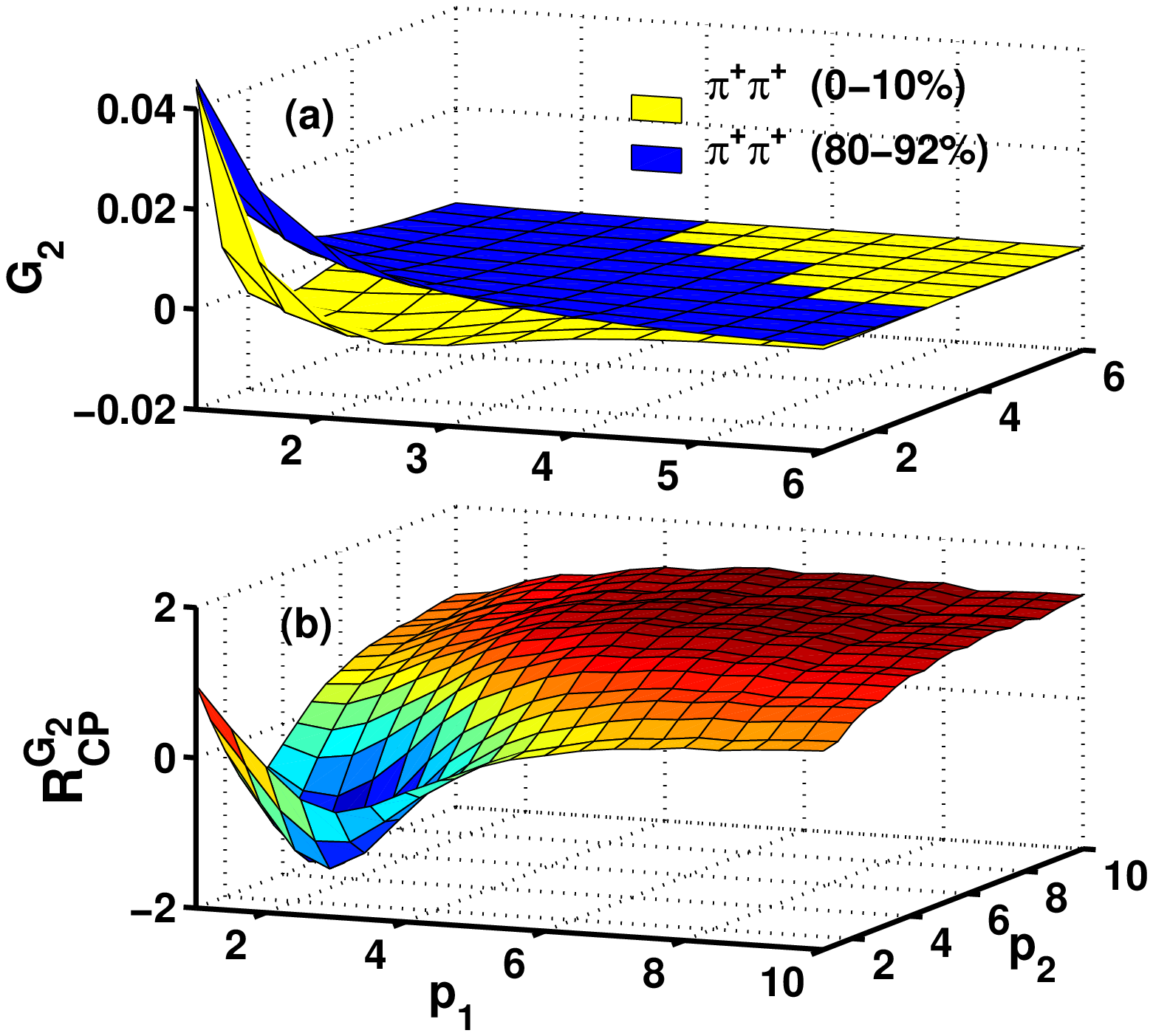}
\caption{(a) $G_2(1,2)$ for two centralities. (b) The ratio $R^{G_2}_{CP}$ of central to peripheral $G_2$.}
\end{minipage}
\hspace*{.3in}
\begin{minipage}[b]{3in}
\includegraphics[width=1.0\textwidth]{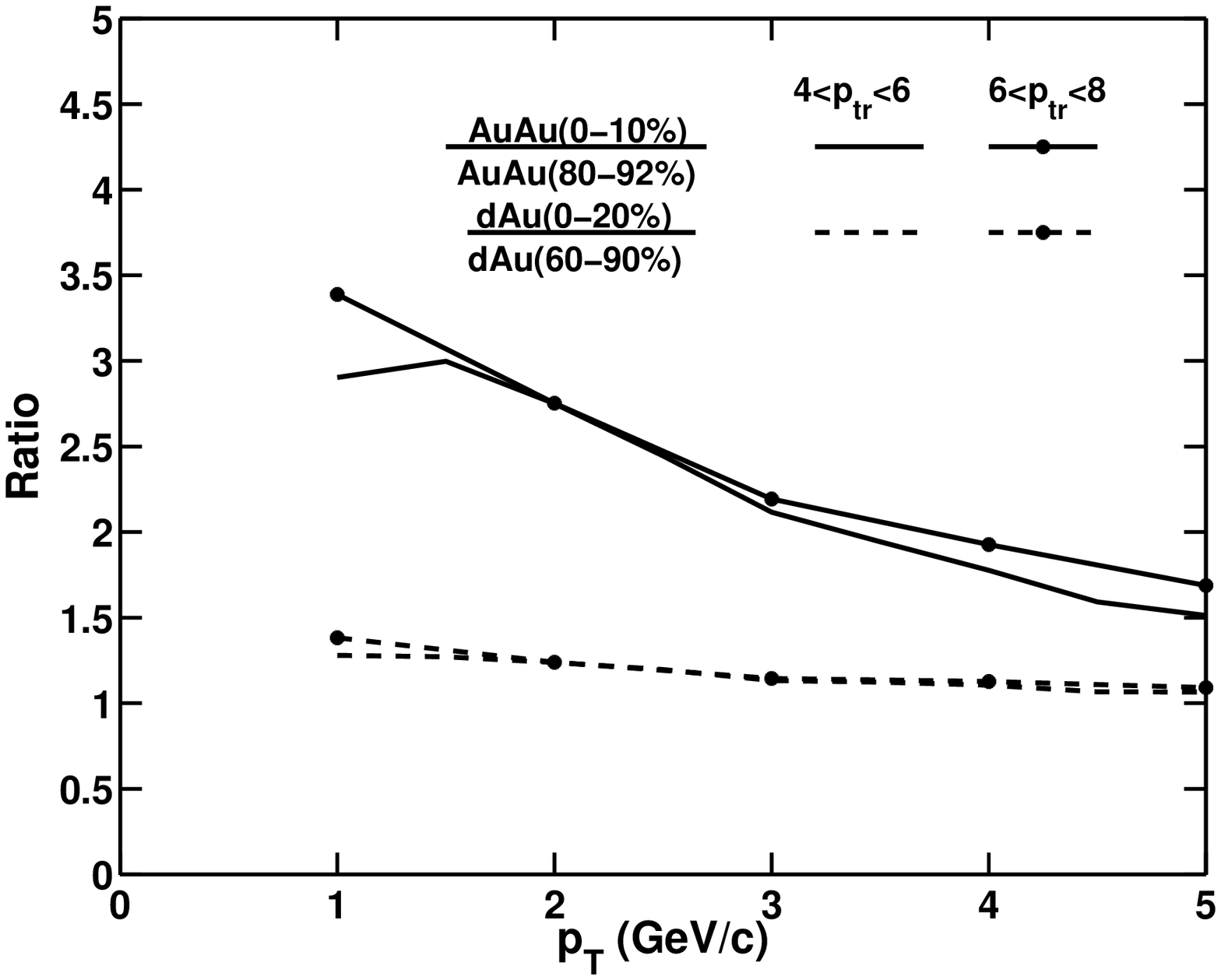}
\caption{$R_{CP}$ vs the momentum of the
AP $(\pi^+)$ for two trigger windows.}
\end{minipage}
\end{figure}

\section{Associated particle distributions}
If we use $p_1$ to denote the transverse momentum of the trigger, and
$p_2$ that of the associated particle, then the per-trigger distribution
of the latter for pions is
\begin{eqnarray} 
{dN^{AP}_{\pi}  \over  p_2 d p_2}= {\int dp_1 p_1\ {dN_{\pi \pi} / 
 p_1 p_2 dp_1d p_2} \over  \int dp_1 p_1\  {dN_{\pi} /
 p_1 dp_1}} \quad ,
\label{rch18}
\end{eqnarray}
where $p_1$ is integrated over a range that corresponds to the
experimental cut on the trigger momentum.  The integrands in the
numerator and denominator are, respectively, (\ref{rch15}) and
(\ref{rch4}).  The associated particle distribution (APD) has been
calculated for both dAu and AuAu collisions at various centralities
\cite{ht2}.  It is found that the $p_T$ distribution of the APD for dAu
collisions has negligible dependence on centrality, although the
dependence is quite significant for AuAu collisions.  Those results are
summarized in Fig.\ 10.  They are different from the results reported in
\cite{hy5} because different quantities are calculated:  whereas
(\ref{rch18}) corresponds to the ratio of integrals, that in \cite{hy5} is
the integral of the ratio.  Comparison with the experimental data
should be made only when the appropriate quantity is chosen that
corresponds to what is measured and analyzed.

Let us now go on to the angular dependence of the APD.  To exhibit
the structures of the near- and away-side jets, it is necessary to make
background subtraction of the data.  F.\ Wang has presented data on
$\Delta \phi$ and $\Delta \eta$ distributions, where the subtraction
scheme used results in the vanishing of the APD in $\Delta \phi$ at
$\left|\Delta \phi \right| = 1$ \cite{fw}.  The corresponding
distribution in $\Delta \eta$ shows a pedestal on top of which sits a
peak at $\Delta \eta= 0$.  To address these features found in the data
it is necessary for us to generalize the formalism that we have used for
recombination.  So far our consideration has only been
one-dimensional (1D), where the parton and hadron momenta are all
collinear.  Now we must consider shower partons in a jet cone that has
3D characteristics.  Furthermore, we must take into account the energy
loss of the hard partons and the subsequent hadronization of the
medium that has absorbed the radiated energy.  These aspects of
generalization have been considered in \cite{ch}.

The fact that the APD in $\Delta \eta$ has a pedestal, not found in the
$\Delta \phi$  distribution, suggests the basic lack of symmetry
between the longitudinal and azimuthal directions.  Indeed, whereas
there is longitudinal expansion of the compressed medium, there is no
azimuthal expansion in the transverse plane, only radial expansion. 
That means that there is no mixing of the various $\phi$ sectors,
making possible the implication that the particles detected in the peak
region with $\left|\Delta \phi \right| < 1$ arise from partons, soft or
hard, that are originally in the same $\phi$ sector, i.e., where the
trigger is measured in the azimuth.  That is not the case with the
$\eta$ variable due to longitudinal expansion, and therein lies the
possibility of a pedestal outside the peak region in $\Delta \eta$
where the trigger is.

Endowing the parton momenta with vectorial properties in 3D, we use
$\psi$ to denote the angle between $\vec{q}_2$ and $\vec{k}$,
assuming for simplicity that $\vec{q}_1$ is along $\vec{k}$, based on
the recognition that it is the relative angle between $\vec{q}_1$ and
$\vec{q}_2$ that matters.  We further assume that there is enough
dispersion of the thermal partons around the average direction at any
$(\eta,\phi)$ such that the hadron momenta $\vec{p}_1$ and
$\vec{p}_2$ due to ${\cal TS}$ recombination can be directed along $\vec{q}_1$ and
$\vec{q}_2$, respectively.  That means that $\psi$ is also the angle
between the measured pion momenta.   Consequently, it is possible to
relate $\psi$ to the pseudorapidities $\eta _1$ and $\eta _2$.  Let us
describe the angular distribution of the shower partons around the  jet
axis by a Gaussian
\begin{eqnarray}
G(\psi,x) = exp \left[- \psi^2/2\sigma^2(x) \right] \quad ,
\label{rch19}
\end{eqnarray}
where the width depends on the momentum fraction $x$ as 
\begin{eqnarray}
\sigma (x) = \sigma_0(1 - x)  \quad ,
\label{rch20}
\end{eqnarray}
which is a simple way to capture the property that the jet cone is
wider for softer partons.  $\sigma_0$ is a free parameter that is to be
determined phenomenologically.

With the angular variable described above we can now write down the
contribution to $F_4$ that gives rise to the trigger at $\vec{p}_1$ and
the AP at $\vec{p}_2$ within the peak in $\Delta \eta$ through ${\cal
TS}$ recombination for both pions
\begin{eqnarray}
F^{TSTS}_4 = \xi \sum_i \int dk k f_i(k) {\cal T}
(q_3)\left\{S(q_1),S(q_2)\right\} {\cal T} (q_4) G (\psi, q_2/k)
\label{rch21}
\end{eqnarray}
where $q_1$ and $q_3$ form the trigger at $p_1$, and $q_2$ and
$q_4$ form the AP at $\vec{p}_2$ at an angle $\psi$ relative to
$\vec{p}_1$.  ${\cal T}(q)$ has the same form as in (\ref{rch6}), but
the inverse slope $T$ is now allowed to be higher than the value used in
the past in order to take into account the enhanced thermal
distribution due to the loss of energy of the hard parton while
traversing the medium.  Since the enhanced thermal partons are in the
immediate vicinity of the hard parton, they are the ones that
recombine with the shower partons, as expressed in (\ref{rch21}).  How
$T$ differs from the value $T_0$ determined from the soft hadron
distribution for $p_T < 2 $ GeV/c without being in the presence of any
jet is another parameter $\Delta T$ in the problem.  We determine 
$\Delta T$ from the pedestal height.  In the recombination model we
are able to attribute the pedestal effect to the difference of ${\cal TT}$
recombination when there is a jet and ${\cal T}_0{\cal T}_0$
recombination of the background in the absence of a jet.  That is, for
the pedestal we have
\begin{eqnarray}
F^{ped}_4 = \xi \sum_i \int dk k f_i(k) S(q_1) {\cal T}
(q_3)\left[{\cal T}(q_2) {\cal T} (q_4)- {\cal T}_0(q_2){\cal T}_0
(q_4)\right]
\label{rch22}
\end{eqnarray}
where the part before the square brackets is for the trigger as in
(\ref{rch21}), while the quantity inside the square brackets is what
remains of the thermal partons for recombination after background
subtraction.  Using the value ${\cal T}_0 = 0.317$ GeV/c we can
determine $T = T_0 + \Delta T$ by varying $\Delta T$ to fit the data
on the pedestal.

The result of our calculation for the APD in $\Delta \eta$ is shown in
Fig.\ 11 \cite{ch}.  The two free parameters $\sigma_0$ and $\Delta T$
are adjusted to get the good fit of the STAR data \cite{fw}; they are
\begin{eqnarray}
\sigma_0 = 0.22, \quad \Delta T = 15 \,  {\rm MeV/c} \quad .
\label{rch23}
\end{eqnarray}
The change from $T_0$ to $T$ is only 5\%, so the difference is
insignificant in the calculation of single-particle distribution. 
However, the difference is sufficient to give rise to a pedestal in
$\Delta \eta$, whose origin is therefore the feedback from the lost
energy of the hard parton to the hadrons through the enhanced
thermal medium.  It should be recognized that the good fit in Fig.\ 11 is
not a trivial consequence of the use of two parameters, since the
height of the peak is the result of multidimensional integrals involving
many components of recombination.  

\begin{figure}
\begin{minipage}[b]{3in}
\includegraphics[width=1.0\textwidth]{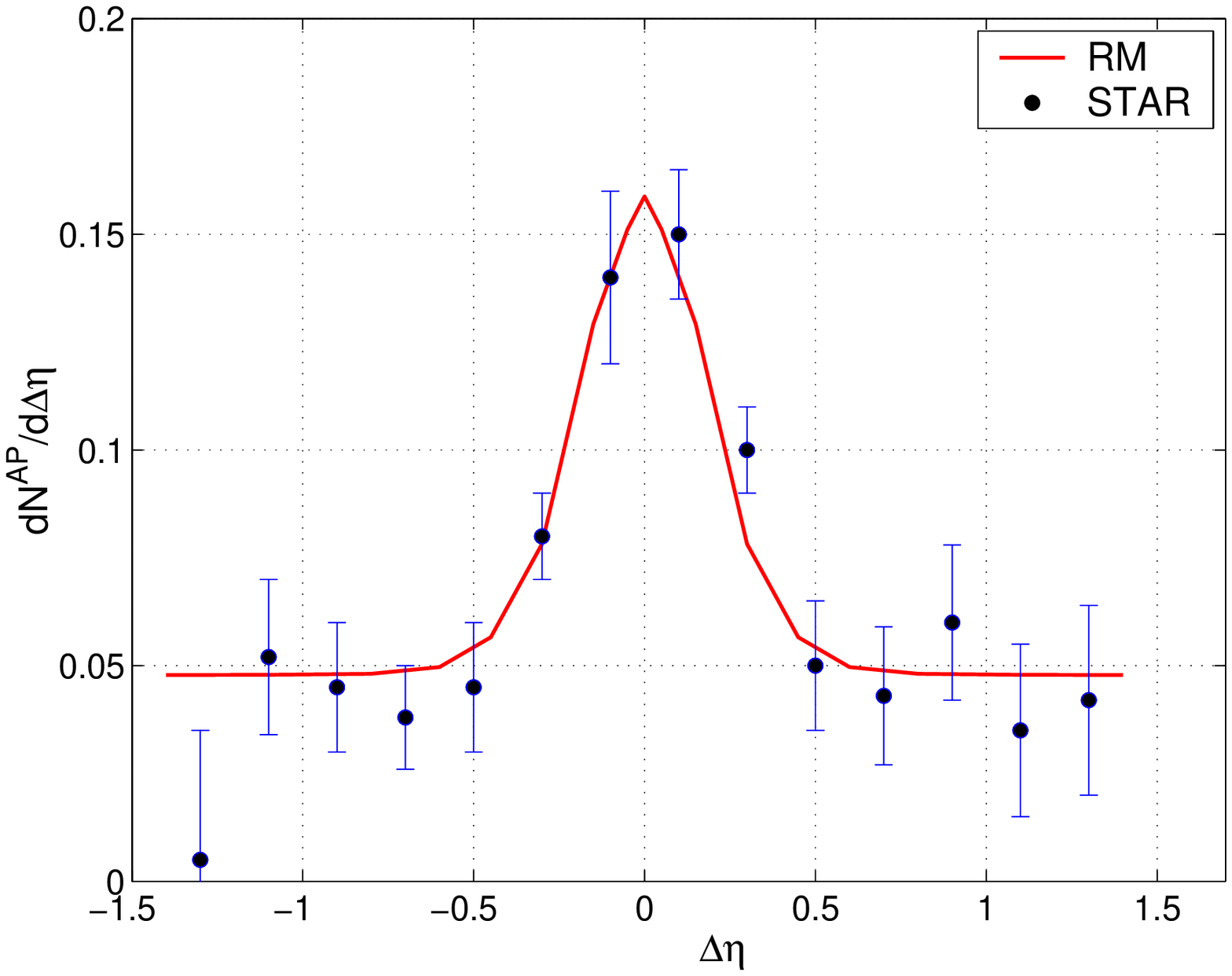}
\caption{APD in $\Delta \eta$ for
$2<p_T<4$ GeV/c with trigger particle in $4<p^{trig}_T<6$ GeV/c. 
The data from \cite{fw} are for all charged hadrons in the respective
$p_T$ ranges.}
\end{minipage}
\hspace*{.3in}
\begin{minipage}[b]{3in}
\includegraphics[width=1.0\textwidth]{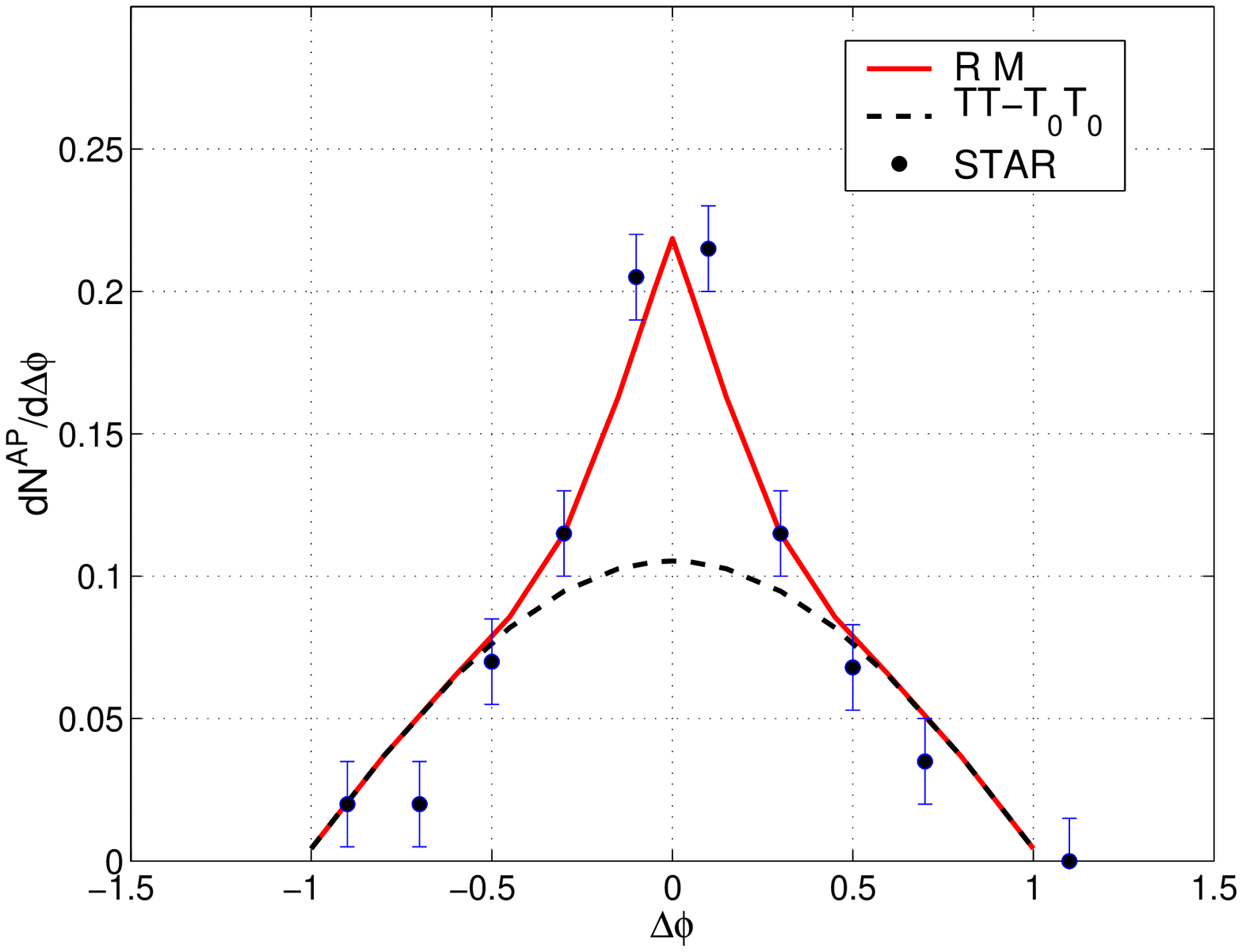}
\vspace*{.1cm}
\caption{Same as in Fig.\ 11 except that the distribution is in $\Delta
\phi$.  The dashed line represents the pedestal effect in $\Delta \phi$.}
\end{minipage}
\end{figure}

With the parameters in (\ref{rch23}) fixed the APD in $\Delta \phi$ can
be calculated; the result is shown in Fig.\ 12 \cite{ch}.  The dashed line
corresponds to the pedestal in $\Delta \eta$, but is forced to vanish at
$\left| \Delta \phi \right| = 1$ because of the subtraction scheme.  The
solid line includes the ${\cal TS}$ contribution on top of the mount
and exhibits good agreement with the data \cite{fw}.  The yield of the
AP has been studied in \cite{fbm} on the basis of some correlation
among the soft partons.

Although we have not put in by hand any short- or long-range
correlation, one may interpret the correlation of the shower partons
exhibited in Fig.\ 5 as an intrinsic short-range correlation in a jet, and
the feedback mechanism of the energy loss of hard partons to the
enhancement of thermal partons as a form of long-range correlation. 
Through parton recombination these correlations are transmitted from
the partons  to the hadrons that are measured.

The advantage of studying the particles associated with triggers is that
the details of the jet structure become manifest and allow us to
determine the properties of the shower partons, such as $\sigma_0$
in (\ref{rch23}).  The drawback is the necessity of making subtraction
of the background that may involve some ambiguity.  A way to avoid
the drawback is to study autocorrelation in $\Delta \eta$ and $\Delta
\phi$ starting with either (\ref{rch11}) or (\ref{rch17}), in which the
two particles are treated on equal footing and no subtraction beyond the
definition of $C_2 (1,2)$ is needed.  That will be our next project, in
which we can use the results of \cite{ch} as the basis for the
calculation of the autocorrelation in $\Delta \eta$ and $\Delta \phi$
within a definite $p_T$ range.

\section{Concluding remarks}
In conclusion, it is worth stressing that parton recombination provides
a framework to describe correlation at intermediate $p_T$ range.  So
far we have not assumed any exotic correlation among the partons,
since none seems necessary.  However, some dynamical correlation
may be present when probed properly, in which case our formalism
may be well suited to decipher its characteristics.  Of course, it is the
correct hadronization process that must first be established.  After that
we can then investigate not only the detail properties of the
correlation in the near-side jet, but also the nature of jet quenching on
the away side.

I am grateful to Gunther Roland and Tom Trainor for inviting me to this very stimulating workshop on Correlations and Fluctuations. I also want to thank my collaborators,  C.\ B.\ Chiu, R.\ Fries, Z.\ Tan and C.\ B.\ Yang, who have been crucial in the completion of the work reviewed here, so the credit for the success of our approach should go mostly to them, without
whom there are no results to show.  This work was supported  in
part,  by the U.\ S.\ Department of Energy under Grant No.
DE-FG02-96ER40972.


\vspace*{1cm}
\begin{thebibliography}{99}

\bibitem{hy1} Hwa R C and  Yang C B 2004 {\it Phys. Rev.} C {\bf 71}
024905

\bibitem{hy2}
Hwa R C and  Yang C B  2004 {\it Phys. Rev.} C
{\bf 70} 024904

\bibitem{hy3} 
Hwa R C and  Yang C B  2002 {\it Phys. Rev.} C
{\bf 66} 025205

\bibitem{sgf} Srivastava D, Gale C and  Fries R 2003 {\it Phys. Rev.} C
{\bf 67} 034903

\bibitem{ht}
 Hwa R C and Tan Z nucl-th/0503052.

\bibitem{ssa1}Adler S (PHENIX Collaboration) 2003 {\it Phys. Rev.
Lett.}  {\bf 91} 072301 

\bibitem{ssa2} Adler S (PHENIX Collaboration) 2004 {\it Phys. Rev.} C {\bf 69} 034909

\bibitem{hy3.5} Hwa R C and  Yang C B  2004 {\it Phys. Rev.} C {\bf 67}
034902


\bibitem{gre}Greco V, Ko C and L\'{e}vai P 2003 {\it Phys.  Rev.
Lett.} {\bf 90}  202302; 2003 {\it Phys.\ Rev.} C {\bf 68}, 034904


\bibitem{rjf} Fries R, M\"{u}ller B, Nonaka C and 
Bass S 2003 {\it Phys.\ Rev.\ Lett.} {\bf 90} 202303 ;  2003 {\it Phys.
Rev.} C {\bf 68} 044902

\bibitem{cr}Cronin J et al  1975 {\it Phys. Rev.} D {\bf 11}
3105;  Antreasyan D et al 1979 {\it Phys.  Rev.} D {\bf 19} 764


\bibitem{hy4} Hwa R C and  Yang C B 2004
{\it Phys. Rev. Lett.} {\bf 93} 082302;  {\it Phys. Rev.} C {\bf 70}, 037901.

\bibitem{fm} Matathias F (for PHENIX Collaboration) 2004 {\it J.
Phys. G:  Nucl. Part. Phys.} {\bf 30} S1113

\bibitem{iar} Arsene I et al (BRAHMS Collaboration) 
2004 {\it Phys. Rev. Lett.} {\bf 93} 242303

\bibitem{hyf}  Hwa R C and  Yang C B  and Fries R J 2005
{\it  Phys. Rev.} C {\bf 71} 024902

\bibitem{rt} See contributions by  Ray L and Trainor T in these
proceedings

\bibitem{ht2}
Hwa R C and Tan Z nucl-th/0503060

\bibitem{hy5} 
Hwa R C and  Yang C B  2004 {\it  Phys. Rev.} C
{\bf 70} 054902

\bibitem{fw}  Wang F (for STAR collaboration) nucl-th/0501016 see
contribution in these proceedings

\bibitem{ch}
Chiu C B and  Hwa R C  nucl-th/0505014

\bibitem{fbm}
Fries R, Bass S and  M\"{u}ller S 2005 {\it Phys.\ Rev.\ Lett.} {\bf 94}
122301

\end{thebibliography}
\end{document}